\documentclass[useAMS,usenatbib]{mn2e}

\usepackage{times}
\usepackage{graphicx}

\newcommand{\kms}   {km~s$^{-1}$}

\newcommand{\eg}    {e.\,g.}
\newcommand{\ie}    {i.\,e.}

\newcommand{\sii}   {[S\,{\sc ii}]}
\newcommand{\nii}   {[N\,{\sc ii}]}

\newcommand{\oi}     {[O\,{\sc i}]}
\newcommand{\oii}    {[O\,{\sc ii}]}
\newcommand{\fe}    {[Fe\,{\sc ii}]}
\newcommand{\vlsr}  {$V_\rmn{LSR}$}
\newcommand{\vtot}  {$V_\rmn{tot}$}
\newcommand{\vtan}  {$V_\rmn{T}$}
\newcommand{\den}    {$n_\mathrm{e}$}
\newcommand{\Te}    {$T_\mathrm{e}$}
\newcommand{\vshock}  {$V_\rmn{bs}$}

\title[Integral Field Spectroscopy of HH 223]
{
INTEGRAL FIELD SPECTROSCOPY OF THE BRIGHTEST KNOTS OF HH~223 IN L723}

\author[R. L\'opez et al.]
{R. L\'opez,$^{1}$\thanks{E-mail:
\mbox{rosario@am.ub.es;
bgarcia@iac.es;robert.estalella@am.ub.es};
\mbox{
angels.riera@upc.edu};
carrasco@mpifr-bonn.mpg.de;
gabriel.gomez@gtc.iac.es
}
B. Garc\'\i a-Lorenzo,$^{2,3}$\footnotemark[1]
R. Estalella,$^{1}$\footnotemark[1]
A. Riera,$^{4}$\footnotemark[1]
\newauthor 
C. Carrasco-Gonz\'alez,$^{5}$\footnotemark[1] 
and
G. G\'omez$^{6,2}$\footnotemark[1]
\thanks{Based on observations collected at the 4.2 m William Herschel and 
2.6 m Nordic
Optical Telescopes at the Observatorio Roque de los Muchachos of 
the Instituto de Astrof\'{\i}sica de Canarias. 
}\\
$^{1}$Departament d'Astronomia i Meteorologia (IEEC-UB), Institut de Ci\`encias
del Cosmos, Universitat de Barcelona, Mart\'{\i} i Franqu\`es 1, 
E-08028 Barcelona, Spain.\\
$^{2}$Instituto de Astrof\'{\i}sica de Canarias, E-38200 La Laguna, Spain.\\
$^{3}$Departamento de Astrof\'{\i}sica, Universidad de La Laguna, E-38205,
 Tenerife, Spain.\\
$^{4}$Dept.\ F\'{\i}sica i Enginyeria Nuclear. EUETI de Barcelona.
Universitat Polit\`ecnica de Catalunya. Comte d'Urgell 187, E-08036 Barcelona,
Spain.\\
$^{5}$Max-Planck-Institut f\"ur Radioastronomie (MPIfR). Auf dem H\"ugel 69, 53121 Bonn,
Germany.\\
$^{6}$GTC Project Office, GRANTECAN S.A. (CALP), E-38712 Bre\~na Baja, La Palma,
Spain.\\
}

\begin{document}

\date{Accepted 2012 May 9. Received 2012 May 9;in original form 2012 March 28}
\pagerange{\pageref{firstpage}--\pageref{lastpage}} \pubyear{55}

\maketitle

\label{firstpage}

\begin{abstract} HH~223 is the optical counterpart of a larger scale H$_2$
outflow, driven by the protostellar source VLA~2A, in L723. 
Its
poorly collimated and rather chaotic morphology suggested the
Integral Field Spectroscopy (IFS)  as an appropriate  option to map the
emission  for deriving  the physical conditions and the
kinematics.  Here we present new results based on the IFS observations 
made with the
INTEGRAL system at the WHT. 
The brightest knots of
HH~223 ($\sim$~16~arcsec, $\simeq$ 0.02 pc at a distance of 300~pc) were mapped 
with a single pointing in
the spectral range 6200--7700 \AA. 
We obtained the emission-line intensity maps for H$\alpha$, \nii\ 6584 \AA\ and \sii\ 6716,
6731 \AA, and
explored the distribution of the excitation and electron density  from
\nii/H$\alpha$, \sii/H$\alpha$, and \sii\ 6716/6731 line-ratio maps. 
Maps of the radial velocity field were obtained.
We analysed the 3D-kinematics  by combining the knot 
radial velocities, derived from IFS data, with the knot proper motions
derived from multi-epoch, narrow-band images.
The intensity maps built from IFS data reproduced well the morphology found
in the
narrow-band images. 
We checked the results obtained from previous long-slit observations with those derived from 
IFS spectra extracted with a similar spatial
sampling. At the positions intersected by the slit, the physical conditions 
and
kinematics derived from IFS are  compatible with those derived 
from long-slit data. In contrast,
significant discrepancies were found when  the 
results from
long-slit data were compared with the  ones derived from IFS spectra 
extracted at positions
shifted a few arcsec from those intersected by the slit. 
This  clearly revealed IFS observations as the best choice to get a
reliable picture of the HH emission  properties.
\end{abstract}

\begin{keywords}
ISM: jets and outflows --
ISM: individual: HH~223, L723, VLA~2
Techniques: Imaging spectroscopy: Integral Field Spectroscopy
\end{keywords}

\section{INTRODUCTION}

The Herbig-Haro object 223 of the Reipurth Catalogue \citep{rei94} is a
knotty, undulating emission of $\sim$~30~arcsec length, and
corresponds to the brightest  optical counterpart of a large-scale 
($\sim$~5.5~arcmin) H$_2$ outflow. The outflow is located in the dark cloud 
Lynds 723 (L723), at a
distance of 300~$\pm$~150~pc (\citealp{gol84}). It is associated with the
low-mass multiple system VLA~2 (\citealp{Ang91}), being one of the components of
this system (VLA~2A) the outflow exciting source (\citealp{Car08}). For a more
detailed description of the region and the nomenclature of the outflow
knots, see \citet{Lop10a} and references therein.

In a previous work (\citealp{Lop09}), we performed long-slit spectroscopy in the
spectral range 5800-8300 \AA\ along two slit positions, which intersect
the peaks of the bright HH 223 knots (A to E), including  
emission from the low-brightness nebula surrounding the knots. The spectra 
of the knots are
characteristic of shock-excited gas with an intermediate/high degree of 
excitation.
The radial velocities derived for the knots are highly blueshifted relative to the ambient
gas cloud, and supersonic, as  expected for a HH optical jet.
Significant changes in the  physical conditions (namely, in the
excitation, ionization and electron density) were found as we move a few arcsec 
along the slits. In addition, 
two velocity components seem to be present at several of the  positions intersected by 
the slits. It is likely that, at these positions, there is contribution to 
the emission from both the
shocked jet gas and the ambient gas, dragged and accelerated by the supersonic
outflow.  
Note, however, that we  sampled partially  the spatial emission of the HH~223  knots
and  interknot nebula by means of the long-slit observations. In contrast, Integral Field Spectroscopy (IFS)
appears to be the appropriate tool to get a wider spatial coverage that includes the whole
spatial knot/interknot emission. This technique has proven to be very 
efficient to study the
kinematics and physical conditions, mainly in the case of Herbig-Haro objects having
a poorly collimated morphology  (see, \eg\ HH~262, \citealp{Lop08};  
HH 110, \citealp{Lop10b}).

Taking advantage of the capabilities of INTEGRAL (\citealp{Arr98}), we made
a single-pointing observation in HH~223, covering  the emission of its brightest knots (A to
D). From these data, we obtained
the morphology in several emission lines, the kinematics and the physical
conditions of HH~223. In addition, we were able to compare the
results derived from IFS data with those obtained from narrow-band and long-slit
observations using similar spatial and spectral resolutions. This experiment gave us
the opportunity to test whether a partial sampling of the emission performed from
long-slit data might mask some behaviours of the actual spatial distribution of the
physical conditions in HHs such as HH~223, with a poor collimated and rather chaotic
morphology.

We also present in this paper the 3D kinematics of the brightest HH~223 knots 
derived by combining the INTEGRAL data (radial velocity) with   narrow-band H$\alpha$
images obtained at three different epochs (proper motions).

\section{OBSERVATIONS AND DATA REDUCTION}

\subsection{INTEGRAL FIELD SPECTROSCOPY}

The 4.2~m William Herschel Telescope (WHT) of the Observatorio del Roque
de los Muchachos (ORM, La Palma, Spain)
 was used in combination with the INTEGRAL fiber optics system (\citealp{Arr98}) and the
 WYFFOS spectrograph (\citealp{Bin94}). INTEGRAL links the Nasmyth focus of the WHT
 with the slit of WYFFOS through three optical fiber bundles. At the focal
 plane, the fibers of these three bundles are arranged into two groups, one
 forming a central rectangle, and the other an outer ring mapping the sky. The
 three bundles can be interchanged online depending on the scientific program or
 the prevailing seeing conditions, as they have different spatial sampling and
 coverage. More information about INTEGRAL bundles and performances can be
 found at the INTEGRAL web page\footnote{http://www.iac.es/proyecto/integral/}. 
 The data
 discussed in this paper were obtained with the standard bundle 2 (STD2). STD2
 consists of 219 fibers (189 science and 30 sky fibers), 
 each of 0.9~arcsec in diameter on the sky and covering 
 a field of view (FOV) of 16~$\times$~12~ arcsec$^2$.
 
\begin{figure*}
\includegraphics[angle=0,width=1\hsize]{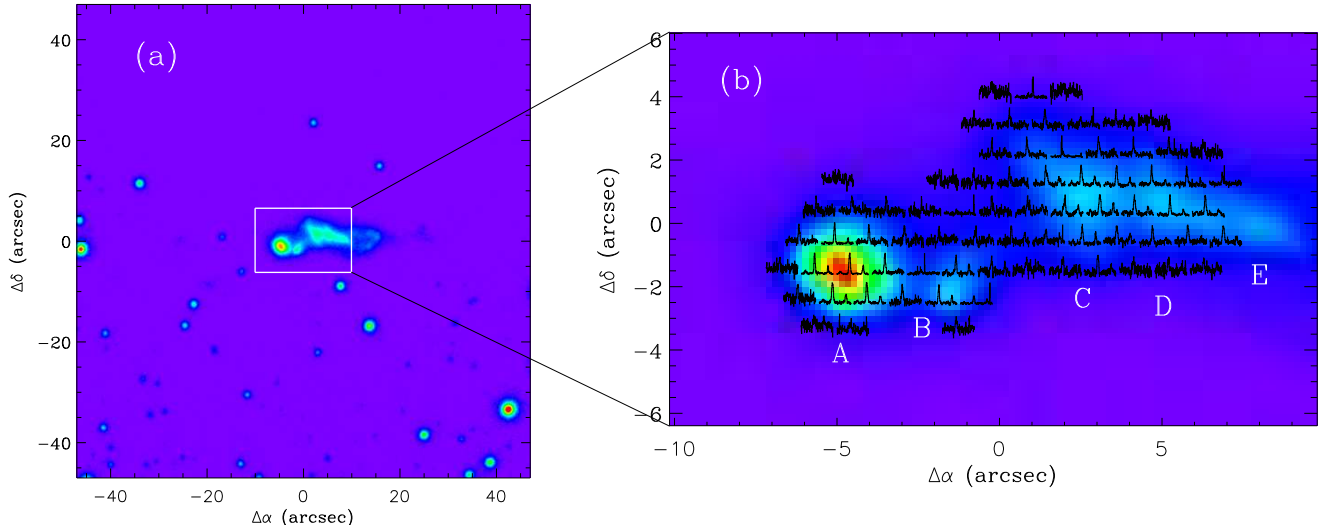}
\caption{(a) H$\alpha$ narrow-band image of the HH~223 field obtained at the
Nordic Optical Telescope (NOT) on 2009 May 18. The white rectangle includes the region
mapped with the IFS single pointing. (b) Two-dimensional distribution diagram of the observed
spectra in the
spectral window 6540-6590 \AA , which includes the H$\alpha$ and \nii\ 6548, 6584 \AA\ emission 
lines. The
spectra have been rescaled to show better  the line profile, and are overplotted
onto the H$\alpha$ narrow-band. Labels of knots identification
(following \citealp{Lop06}) are included. The offsets are relative to the centre
of the field covered by IFS.
\label{cobertura}}
\end{figure*}

The observations were made on 2009 May 25 as a backup program of the science
verification allocated time for a new equalized fiber bundle for INTEGRAL. 
The WYFFOS spectrograph was equipped with a 1200 groove mm$^{-1}$ grating
(centered on $\lambda\ 7000$ \AA ) and a EEV mosaic (two EEV-42-80 thinned and AR
coated CCDs) of 4308$\times$~4200 pixels of 13.5 $\mu$m in size, giving a linear dispersion
of about 0.4 \AA\ pixel$^{-1}$ in the spectral range $\sim$ 6180--7700 \AA . 
With this configuration, and pointing to the brightest knots of HH~223 
($\alpha$=19$^{\rm h}$~17$^{\rm m}$~57$\fs7$,
$\delta$=+19$\degr$~11$\arcmin$~52$\arcsec$), 
five exposures of 1800~s each and one exposure of 900~s were taken 
to get a total effective integration time of 2.75~hr.
This single pointing covers the brightest HH~223 knots 
(HH~223-A to -D) and the low-brightness nebular
emission surrounding these knots, as illustrated in Fig. \ref{cobertura}, where
the spectrum  extracted from each fiber, in a  wavelength range around 
H$\alpha$, is overplotted on the HH~223 H$\alpha$ narrow-band image. The 
knots are labeled as in \citet{Lop06}.

The data were reduced using IRAF\footnote{IRAF is
distributed by the National Optical Astronomy Observatories, which are
operated by the Association of Universities for Research in Astronomy,
Inc., under cooperative agreement with the National Science Foundation.}
standard routines (see \eg\ \citealp{Gar05} 
for a description of the reduction of IFS data from fiber-based instruments). We obtained typical
wavelength calibration errors of 0.1 \AA , which give velocity uncertainties of about 
$\pm$~4.5 km~s$^{-1}$ for H$\alpha$. The differential atmospheric refraction was estimated 
according to the
model given by \citet{All73}, being this effect negligible for the HH~223 observations. 
The night was photometric according to the Mercator Telescope measurements\footnote{
http://www.ast.cam.ac.uk/dwe/SRF/camc-extinction.html} 
and the seeing was around 1 arcsec.

\subsection{Narrow-band CCD images}

Deep narrow-band CCD images of  L723, covering a field of 
$\sim~5~\times$~5~arcmin$^2$ that includes  
HH~223 were obtained at three different epochs (2004 July 20, 2007 July 14, 
and 2009 May 18), suitable to
measure proper motion displacements of the knots. The Andalucia Faint Object Spectrograph and Camera
(ALFOSC) was used on the 2.6~m Nordic Optical Telescope (NOT) at the ORM. The image scale was 0.188 
arcsec~pixel$^{-1}$. The filter used had a central wavelength of 6564 \AA\ and a bandpass of 33 
\AA, which mainly included emission from the H$\alpha$ line, with some 
contamination from the \nii\ $\lambda$ 6548 \AA\ 
emission line. All the images were processed as described in \citet{Lop06} 
for the 2004 observing run. The IAA  
Guaranteed Observing Time with ALFOSC was used for the observing run of 2009.\\

\section{IFS RESULTS}

\subsection{IFS imaging: spatial distribution of physical conditions}

Emissions from H$\alpha$, \nii\ and \sii\ $\lambda$ 6716, 6731 \AA\ were detected
in all the fibers covering the HH~223 knots and its surrounding nebula mapped with
INTEGRAL. Other characteristic HH emission
lines included in the observed wavelength range (\oi\ doublet, \fe\ $\lambda$
7155 \AA\ and \oii\ $\lambda$ 7300 \AA\ doublet) were only detected in a few
fibers, located around the peak positions of the brightest knots A and B. 

\begin{figure*}
\includegraphics[angle=-90,width=0.8\hsize]{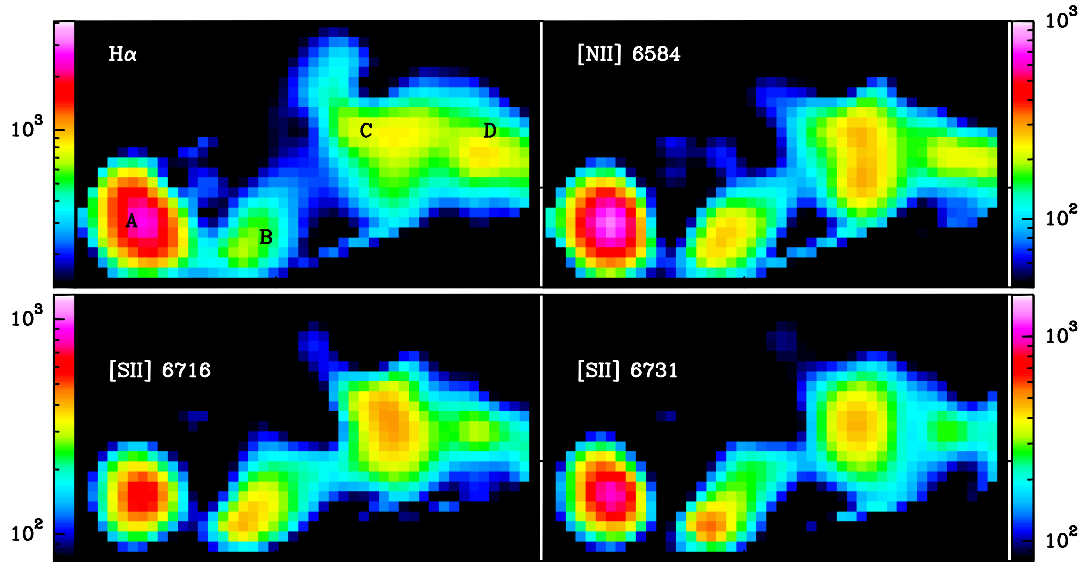}
\caption{
IFS-derived narrow-band images for several emission lines (H$\alpha$, \nii 
$\lambda$ 6584
\AA\ and \sii $\lambda$ 6716, 6731 \AA).  For each position, the flux of the line 
was obtained from
a Gaussian fit to the line profile, after subtracting a continuum, obtained
from the adjacent wavelength range free of line emission. The flux scale is in
counts.
\label{flux}}
\end{figure*}

Figure \ref{flux} shows the morphology of HH~223  for the H$\alpha$,
\nii\ $\lambda$~6584 \AA\ and \sii\ $\lambda$~6716, 6731 \AA\ emission lines
 obtained from IFS data. To
generate these narrow-band images, a two-dimensional interpolation was applied
using the IDA package (\citealp{Gar02}). In particular, an ASCII file with the actual
position of the fiber and the corresponding spectral feature was transformed to obtain a
regularly spaced rectangular grid. In this way, we built up narrow-band images with a field
of view  of 47$\times$27 pixels ($\sim$~14 arcsec~$\times$~8 arcsec) with a spatial
scale of 0.3 arcsec pixel$^{-1}$ that can be treated with  standard astronomical
software packages. 

The IFS maps (Fig.\ \ref{flux}) are in good agreement with previous CCD
narrow-band H$\alpha$ and \sii\ images of HH~223 (\citealp{Lop06}). In the four emission lines considered,
the IFS maps
reproduce the  undulating, knotty structure of the emission, surrounded by a  
low-brightness nebula that had already been found in the narrow-band, CCD images. Note however that the
IFS data allowed us to map separately the emission coming  from the  H$\alpha$ and \nii\ lines,
as well as the two lines of the \sii\ doublet.  We can now observe the close 
similarity between the brightness distribution of  
the \sii\ and \nii\ maps. 
In contrast, the comparison of H$\alpha$ and
\nii\ IFS maps  shows differences  in the
morphology and relative brightness of  the knots.  
In particular, knot C shows a  somewhat different morphology in \nii\ (with two,
north-south,  condensations of similar  brightness) and H$\alpha$ (where the southern
condensation is  fainter than the northern one). Furthermore, knots B and C 
show a higher \nii/H$\alpha$ brightness ratio than knot A.
As we will discuss later, 
these
signatures are the result  of differences in gas excitation through  
the HH~223 emission,  which were only
outlined from long-slit spectroscopy.

\begin{figure}
\centering
\includegraphics[angle=0,width=0.95\hsize]{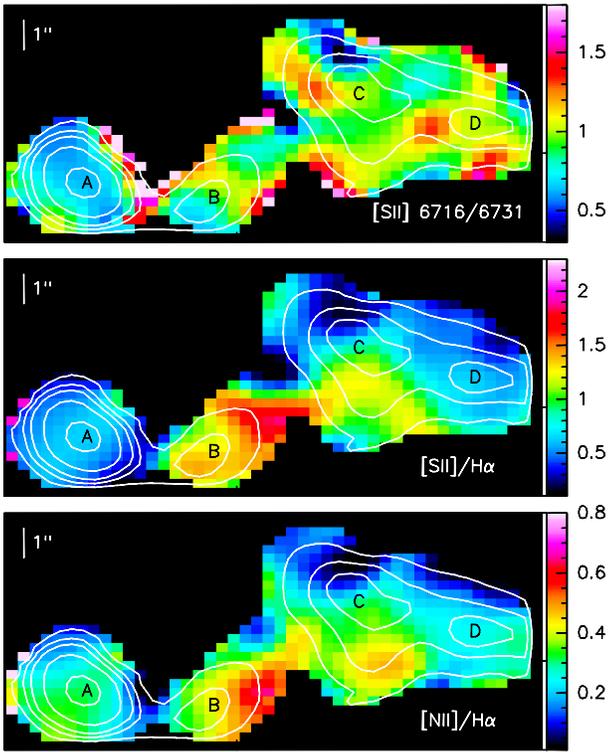}
\caption{
 \sii 6716/6731 (top), \sii\ (6716+6731)/H$\alpha$ (middle) and \nii\ 6584/H$\alpha$ (bottom)
 line-ratio maps obtained
from the IFS-narrow-band images, blanking the regions with a SNR $\leq$ 3. Contours from the
H$\alpha$ IFS image are overlapped to help with the identification of the knot condensations.
\label{lineratio}}
\end{figure}

Line-ratio maps of the emission  lines widely used for
nebular diagnosis (\sii\ (6716+6731)/H$\alpha$ and \nii\ 6584/H$\alpha$ for excitation,  
and \sii\ 6716/6731 for electron density) were obtained from the
IFS-narrow-band images, and are displayed in  Fig. \ref{lineratio}.

For all the mapped region, which included emission from knots and from the
low-brightness
interknot nebula, the values of the \sii/H$\alpha$ line ratio  range from 0.6 to 1.4, and
those of the \nii/H$\alpha$, from 0.2 to 0.5. According to the  \sii/H$\alpha$  vs.
\nii/H$\alpha$ diagnostic diagrams  (see \eg\ \citealp{Can81}), these values are found in
HH objects, where the emission is produced by shocks. In fact, the
values derived for the  \sii/H$\alpha$ line-ratio correspond to the spectra with an
intermediate/high degree of excitation, following the classification for HHs 
proposed by \citet{Rag96}, which
distinguished the low (with a \sii/H$\alpha$ $\ge$ 1.5) from the intermediate/high excitation 
spectra. The spatial distribution of the electron density (\den) is outlined in the 
\sii~6716/6731 line ratio map. Values of the sulphur line ratio range from 0.6 to 1.2,
which correspond to  \den\ ranging from 0.24 to 4.2$\times$10$^3$ cm$^{-3}$ (\den\ derived
using the {\small TEMDEN} task of the {\small IRAF/STSDAS} package, and assuming \Te=10$^4$
K).

A previous work (\citealp{Lop09}) partially sampled the physical conditions  in
HH~223 by means of long-slit observations, with a slit positioned along HH~223 crossing the
knot intensity peaks. The long-slit  data revealed that knot A had the highest excitation
spectrum. A trend of increasing excitation going towards the west was
found for the
rest of the knots. The \den\ values derived for knots A and B were significant 
higher (by factors of $\sim$ 5 to 10) than for the rest of the knots. Hence,
 the line-ratio values
found for the  positions observed in the long-slit HH~223 sampling are consistent
with those obtained, at these positions, from the line-ratio IFS maps. Note, however, that
the IFS data are better suited to map the physical conditions in the whole HH~223 region
covered  by the single IFS pointing.  As a result, the maps of Fig.~\ref{lineratio} show a rather
more complex spatial distribution for both, excitation and density, which 
was not self-evident
from the partial sampling of long-slit data. 

The electron density (\den), derived from the \sii\ 6716/6731 map, is 
higher in the southern, eastern knots (A, B), with \den\ $\sim$
4200-2700 cm$^{-3}$ respectively, than in the other two mapped knots (C, D), 
with \den\ $\sim$
650-500~cm$^{-3}$. In addition, the \den\ map has revealed other density enhancements
offset from
 the knots intensity  peaks (\eg\ the one  
towards the northwest of knot B, within the low-brightness nebula connecting knots B
and C, with
\den\ $\sim$ 1400~cm$^{-3}$), as well as a variation of \den\ across knot B,
moving from the northwest to the southeast, and ranging  from $\sim$ 850 to $\sim$
2700~cm$^{-3}$. 
Hence, the  \sii\ 6716/6731 map  likely suggests density inhomogeneities within the
emitting gas at smaller  scales than the knot sizes.

 The spatial distribution of the gas excitation is outlined in the  \sii/H$\alpha$ and
 \nii/H$\alpha$ maps of Fig.~\ref{lineratio}. The emission with the highest degree of 
 excitation is found for
 knot A, and also around the peak intensity of knot C, while the emission associated
 with knot B has the lowest excitation degree.  However, like for the spatial distribution of the
 electron density, the overall pattern is very complex, suggesting again smaller-scale inhomogeneities
 in the excitation conditions within the mapped region.

\subsection{IFS radial velocity field}

\begin{figure*}
\centering
\includegraphics[angle=-90,width=0.65\vsize]{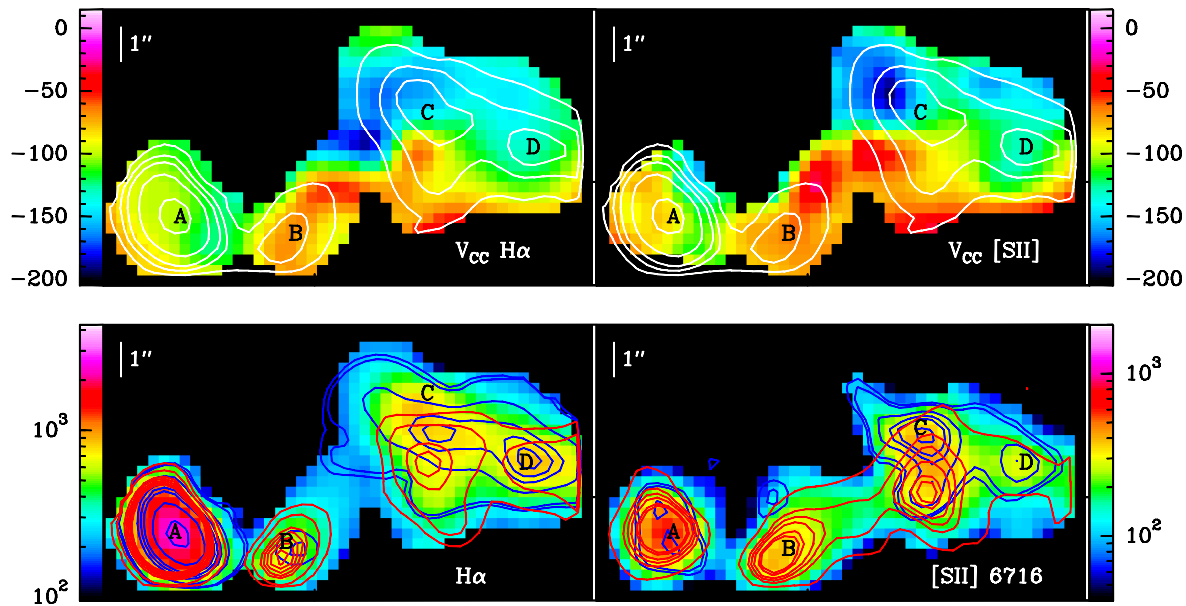}
\caption{Lower panels: Integrated intensity maps obtained by integrating the
signal within all the  H$\alpha$ (left) and \sii\ $\lambda$ 6716 \AA\ 
(right) line profiles (zero-order intensity momenta). 
For each of the  panels, the red contours correspond
to the emission intensity obtained by integrating the signal for \vlsr\ $\geq$ --90
 \kms (LVC) and the blue contours,  to the emission intensity for  \vlsr\ 
 $\leq$ --90 \kms.
Upper panels: Radial velocity field obtained  from cross-correlation 
in the  H$\alpha$+\nii\ range (left), and \sii\ 6716, 6731 \AA\ ranges 
(right). The  velocity scale is in \kms. White contours, as in
Fig.~\ref{lineratio}.
\label{vral}}
\end{figure*}

Radial velocity maps were first derived from the line centroids of a
single-Gaussian fit to the H$\alpha$ and \sii\ emission line profiles. However,
a careful inspection of the spectra extracted at several positions on HH~223 
reveals that the emission lines display broad, asymmetric and even double-peaked
shapes, suggesting that at least two 
kinematic components are contributing to
the line profile at these positions. Hence, we also  derived 
the  radial velocities
by applying the cross-correlation technique through the XCORR routine of the
DIPSO package. Ideally, the template must be a high-S/N spectrum, with a
resolution similar to or better than the target spectrum. Unfortunately, we lack
such an ideal template, and we adopted as template the spectrum extracted at the
peak of the brightest knot (knot A),  which seems to have  symmetric line
profiles.
The
velocity maps derived from these two methods are in good agreement and  show a
rather complex pattern in the observed lines. Figure \ref{vral} (upper panel) shows 
the maps of the 
radial velocity field for  the H$\alpha$ and \sii\ emissions. In order to estimate the
radial-velocity uncertainties, we compared the values derived from the two
techniques (single-Gaussian fit to the profiles,  and cross-correlation)  giving
a mean difference in radial velocities of 2$\pm$6 \kms.  The largest differences
between the radial velocity maps derived from the two techniques are found at
positions with clear double-peaked line profiles. 

The derived velocities\footnote{Velocities are referred to the local
standard of rest (LSR) frame. A \vlsr=+10.9~\kms\ for the parent cloud has been
taken from  \citealp{tor86}.} are highly blueshifted, with
values ranging from  --180 to  --60~\kms.
The velocity field shows a rather
complex pattern. In spite of that, some trends can be observed in   
the spatial
distribution of the velocities:
the lowest (absolute) velocity values, ranging from 
 --80 to  --60~\kms, come from the low-brightness emission
that connects knot A with knots C and D, and
from a shell at the south  of the knots A to D intensity peaks. The highest (absolute)
velocity values, ranging from 
 --180 to  --150~\kms, come 
from a region at the
northeast  of knot C intensity peak. Finally, 
we derived velocities 
ranging from $\simeq$ --130 to $\simeq$ --90~\kms\ for the emissions arising around 
the knot A, C and D intensity peaks. 

The line profile shapes of most of the spectra suggested a
contribution to the emission from  two 
 kinematic components. 
To explore
this possibility, we built two chanel maps, by considering
two ranges for the velocity: the higher blueshifted component
(HVC),  for \vlsr$\leq$ --90\kms, and the lower blueshifted component (LVC), for 
\vlsr$\geq$ --90\kms. Then, we obtained the  emission maps  
of the LVC and  HVC, for the  H$\alpha$ and \sii\  lines, 
by integrating the signal within the corresponding velocity range. 
Figure~\ref{vral} (lower  panels) displays the spatial distribution of the two velocity
components. Contours for the HVC (blue) and LVC (red) have been overlapped
onto the
integrated emission (zero-order intensity moment). All the knots have emission
from both kinematic components. However, the
spatial distribution of these two velocity components does not fully coincide
in H$\alpha$ neither in \sii\ lines. The most remarkable
differences appear for knots B and C, where the peak
intensity positions of the HVC and LVC appear shifted. We measured offset
values of 0.9 and 1.5 arcsec in H$\alpha$ and \sii\ respectively for knot C, and
offsets of 0.7 and 1.5 arcsec for knot B. These offset values are reliable, according to the
spatial resolution of the data.

\begin{figure} 
\centering
\includegraphics[angle=0,width=1.0\hsize]{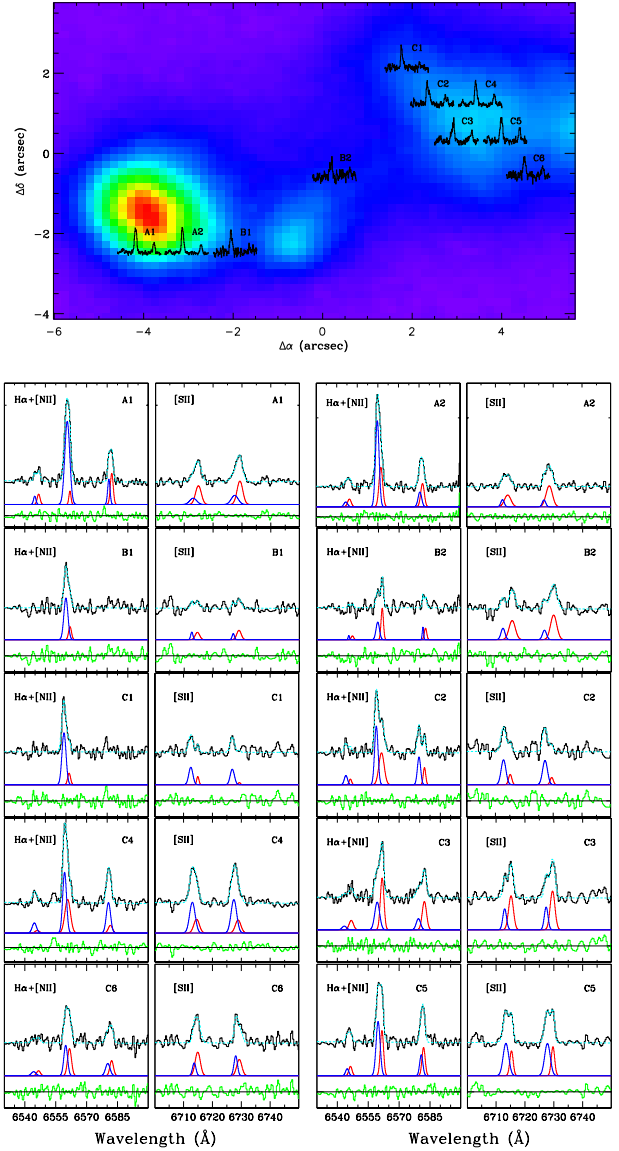}
\caption{Top: Location of the spectra where double-peaked emission lines
are detected, overplotted onto the H$\alpha$ narrow-band image of HH~223 (same as
Fig.\ref{cobertura}).
Bottom: Observed spectra (black, continuous) and fits (blue, dashed) obtained
from the model of two Gaussian components applied to the H$\alpha$+\nii\
6548,6584 \AA\ (left) and \sii\ 6716,6731 \AA\ (right). The two individual
components are shown below in each panel. The same colour (red and blue)
indicates the same component for each line. Residual obtained subtracting the
fit from the observed spectra are plotted in green. Labels in the right corner
of each plot indicate the corresponding spectra plotted in the top panel of this
figure.
\label{specdouble}}
\end{figure}

Clear doubled-peaked emission lines were only found in
a set of spectra  arising from the bright nebula at several positions south of 
 the knot A and B
intensity peak,  and from the region covering knot C (see Fig.~\ref{specdouble}, upper
panel).  We additionally performed a line-profile decomposition
to reproduce the H$\alpha$+\nii\ 6548,6584 \AA, and the \sii\ 6716,6731 \AA\
line profiles of the spectra coming from these regions. The line-profile
decomposition was obtained from a model including two Gaussian components. The
fitting was performed using the DIPSO package of STARLINK, and the results are
shown in Fig.~\ref{specdouble}. The two velocity components derived in this
way appear
blueshifted. For the higher blueshifted  component (HVC), velocities ranging
from --180~\kms\ to --130~\kms\ are obtained.  The velocities derived for  the
lower blueshifted component (LVC) range from --80~\kms\ to --40~\kms. 

Unfortunately, there are few
spectra  where a reliable line-profile decomposition can be obtained, not 
allowing us 
to infer  whether the two kinematic components  are tracing
emission  with different physical conditions, having thus different origin or
not.
However, a trend is found that consists in  the HVC 
having a higher degree of excitation (with  \sii/H$\alpha$ ratios $\leq$
1) and lower electron density (\den\ ranging from $\simeq$ 200 to 800 cm$^{-3}$)
than the LVC, which is less excited
(\sii/H$\alpha$ $\geq$ 1)  and denser (\den\ ranging from 1500 to 6000
cm$^{-3}$). It should be mentioned that a similar behaviour for the excitation
and electron density have been detected in jets emerging from Class 0/I 
 sources and in
microjets from TTauri stars (see, \eg\ \citealp{Ham94}; \citealp{Hir97};
\citealp{Glo08};  \citealp{Glo10}). From the analysis of the double-peaked
line profiles, these authors found that the  emission from the LVC
 usually  presents higher electron density and lower excitation than
the emission coming from the HVC. In most  cases, the
LVC is spatially more compact, and is detected only up at a few arcsec near 
the exciting jet
source, around the jet launching region, being related to the launching
mechanisms. This scenario does not fit our case, because the  knots mapped
in HH~223 are far away ($\sim$ 0.1~pc) from the  outflow exciting source. 
 Two-peaked line profiles have also been observed in the brighter knots of
HH~32 (see \eg\ \citealp{Sol86}, and \citealp{Bec04}). In this case, 
the profiles have been
interpreted as originated from the emission of bow-shocks.
However, at the light of the 3D kinematics derived from our observations, it
is not clear that the bow-shock interpretation could work so well in HH~223 (see
later).
In other  cases 
(\eg\ HH~34
and HH~46-47) the origin of the LVC  at  larger distances from the source is
attributed to the gas entrained and accelerated by the high-velocity outflow.
This might be a plausible 
explanation for the origin of the LVC emission in HH~223. However, a more close 
similarity in the 
spatial distribution of the two velocity components should be expected in this
case. Alternatively, the detection of two velocity components that also differ
in their excitation has been attributed to episodic velocity variations in the
outflow (\eg\ HH~151, \citealp{Mov12}). This possibility (\ie\ episodic
mass-ejection events with variable velocity and direction, as can be expected
when the exciting outflow source belongs to a multiple system) might be a
more plausible explanation 
for the origin of the LVC and HVC  emissions in HH~223.

\subsection{Long-slit vs. IFS}

In spite of the comparisons not being fully straightforward, one of the goals of
these observations was to check whether the results derived from IFS data are fully
consistent with the ones
obtained from the long-slit sampling. Apparently some 
discrepancies  in the spatial behaviour of the
velocity appeared when  IFS results (this work) are compared  with those obtained 
from long-slit
observations of \citet{Lop09}. For example, as can be seen from Fig.~\ref{vral},
low negative
velocity values are found  associated with knot B emission  (\vlsr $\simeq$ --70~\kms
around the H$\alpha$ knot intensity peak), and  high negative velocity values
appear associated with knot C (with \vlsr $\simeq$ --150~\kms\
around the H$\alpha$ knot intensity peak), reaching even higher negative values 
(\vlsr $\leq$ --180~\kms) at positions  north to the knot C intensity peak.  In
contrast, we found velocities  of $\simeq$ --130~\kms and $\simeq$ --70~\kms\
at  positions sampled from long-slit spectroscopy through knots B and C,
respectively. However, we 
concluded that, in this case, the differences found
between the results from IFS and long-slit observations   mainly arise as a result
of  an
incomplete sampling of the highly complex kinematics of HH~223.

\begin{table}
\caption{\label{vtot} Proper motions (\vtan)$^{1}$ and full spatial velocities (\vtot) of HH~223 knots} 
\begin{tabular}{ccccc}
\hline\hline
Knot &  \vtan& PA & \vtot & $\phi$\\
     & (\kms)&(degrees)&(\kms)&(degrees)\\
\hline 
A &$36.4\pm 2.2$&$144\pm6$\phantom{2} &$107.6\pm 5.0$&$70\pm 5\phantom{1}$\\ 
B &$49.2\pm 3.9$&$106\pm 3$\phantom{2}&$84.4\phantom{1}\pm 6.4$&$54\pm
6\phantom{1}$\\ 
C &\phantom{1}$3.0 \pm 6.5$&\phantom{0}$52\phantom{1}\pm 104$&\phantom{0}$120.7\pm
8.3$\phantom{1}&$89\pm 19$\\
D &$15.0 \pm2.4$&$99\phantom{1}\pm6$\phantom{2} &\phantom{0}$120.0\pm
8.7$\phantom{1}&$89\pm 10$\\
E &$60.3\pm 2.7$&$87\phantom{1}\pm 1\phantom{2} $&&\\ 
F1&$40.8\pm 3.9$&\phantom{1}$80\phantom{1}\pm 11\phantom{2} $&&\\ 
F2&$64.3\pm 2.5$&$91\phantom{1}\pm 6$\phantom{2}&&\\ 
F3&\phantom{1}$7.5 \pm 4.2$&\phantom{0}$194\pm 57$\phantom{1}&&\\ 
\hline 
\end{tabular}
\begin{list}{}
\item
$^{1}$ A distance of 300~pc has been adopted. The uncertainty in the L723
distance was not taken into account for derive the errors in proper motions
quoted in the table. 
\end{list}
\end{table}

\begin{figure} 
\centering
\includegraphics[angle=0,width=1\hsize,clip]{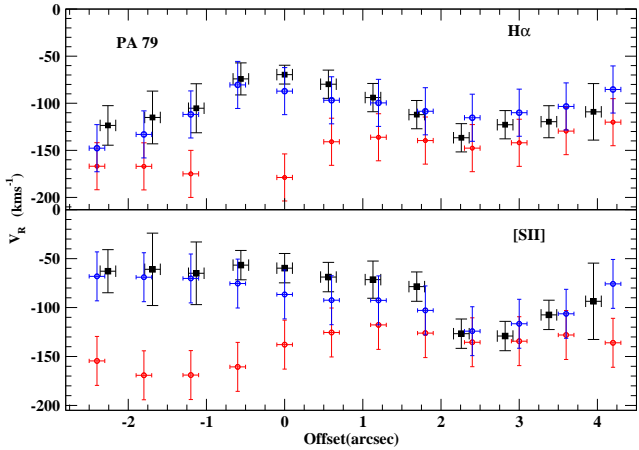}
\caption{Comparison of radial velocities derived from long-slit spectra (black)
and
from IFS spectra, obtained through  simulated slits of 1 arcsec width
positioned at P.A. 79$^\circ$: crossing the knot C intensity peak (blue); the 
same, but 
2 arcsec north of it (red). Offsets are relative to the knot C intensity peak. 
\label{vs79}}
\end{figure}

\begin{figure} 
\centering
\includegraphics[angle=0,width=1\hsize,clip]{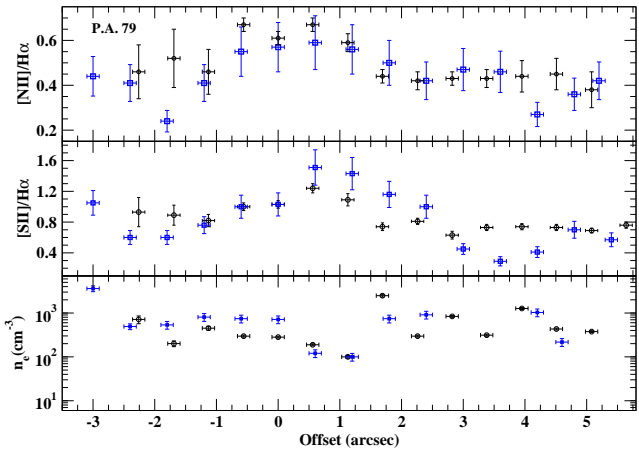}
\caption{Comparison of line ratios obtained from long-slit spectra (black)
and from IFS spectra obtained through  simulated slits of 1 arcsec width and
positioned at P.A. 79$^{\circ}$ crossing the knot C intensity peak (blue).  
\label{cratios2}}
\end{figure}

To make more meaningful the comparisons, we proceeded to simulate
long-slit spectra from the IFS data with a close similar sampling as in long-slit
observations. First, we built a grid of 0.3 arcsec spacing; then, we interpolated
the signal to  extract one-dimensional spectra at positions intersected by two
simulated slits of 1 arcsec width. The simulated 
slits were positioned  crossing the peak intensities of
knots A and C, at position angles (PA) 73$^\circ$ and 79$^\circ$
respectively. 
We derived the
radial velocities (from the centroids) and line ratios (from the fluxes) of the
HH~223 emission within the simulated slits from Gaussian fits to the lines of
these extracted spectra. Figure~\ref{vs79} shows the radial velocities
derived from the centroids of the H$\alpha$ and \sii\ 6716 \AA\ emission lines
(blue). The corresponding radial velocities derived from long-slit spectra are 
represented by the
black dots.  Figure \ref{cratios2} shows the \nii/H$\alpha$ and \sii/H$\alpha$
line ratios (both are tracers of the excitation conditions) and the electron
density, \den, from the \sii\ 6716/6731 line ratio, obtained from the spectra
along the simulated slit (blue) and the corresponding values from long-slit
observations (black). As can be seen from these figures, both the kinematics and
the
physical conditions of the emission intersected by the simulated slit are 
compatible with those found from the long-slit sampling. 
In contrast, some
discrepancies appear when  velocities derived from long-slit are compared
 with those
obtained from IFS spectra through the simulated slit, but
displaced, in parallel, 
$\sim$ 2 arcsec north to the one crossing the knots peaks (red dots in
Fig.~\ref{vs79}).
This test  
reveals the importance of obtaining 2D maps covering the
full spatial extension of the emission, instead of doing 1D long-slit sampling,
mostly in the cases where the kinematics and physical conditions of the emission
show such as complex pattern, as in HH~223. 

\begin{table*}
\caption{\label{vrmp}\vlsr\ (\kms) line centroids$^{1}$ of the
HH~223 knots from IFS data.} 
\begin{tabular}{cccccccc}
\hline\hline
Knot& \nii\ & H$\alpha$& \nii\ & \sii\ & \sii\ & Average & rms \\
    & 6548~\AA  &          &  6583~\AA &  6716~\AA & 6731~\AA  &  &\\
\hline 
A&\phantom{1}$-88.3$&$-101.3$&\phantom{1}$-86.9$&\phantom{1}$-86.1$&\phantom{1}$-85.4$&\phantom{1}$-89.6$&
6.6\phantom{1}\\ 
B
&\phantom{1}$-78.7$&\phantom{1}$-68.6$&\phantom{1}$-77.3$&\phantom{1}$-67.8$&\phantom{1}$-67.2$&\phantom{1}$-71.9$&
5.6\phantom{1}\\ 
C &\phantom{1}$-94.2$&$-120.7$&\phantom{1}$-91.9$&$-103.0$&$-102.3$&$-102.5$&
11.3 \\
D &$-116.7$&$-124.3$&$-115.2$&$-120.7$&$-120.0$
&$-119.4$&3.6\phantom{1}\\ 
\hline 
\end{tabular}
\begin{list}{}
\item
$^{1}$ Typical value of the error is 5~\kms
\end{list}
\end{table*}

\begin{figure} 
\centering
\includegraphics[angle=-90,width=1\hsize,clip]{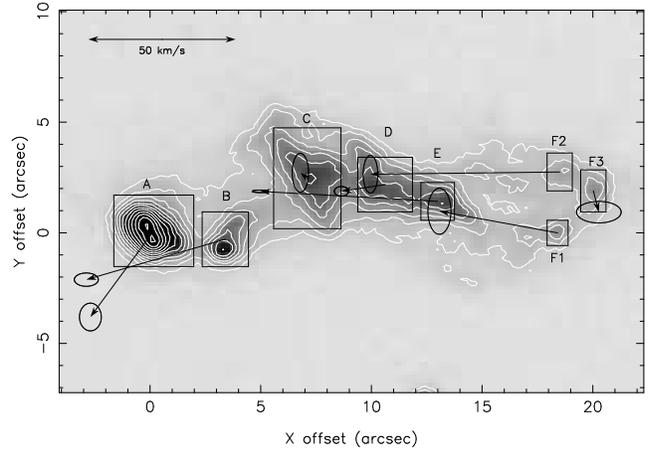}
\caption{CCD H$\alpha$ image of HH~223 from the last epoch (2009). Arrows indicate
the proper motion of the knot. The ellipse at the end of the arrow indicates the
uncertainty in the components of the velocity vector. Boxes include the regions used
for calculation of the knot proper motion. The same boxes have been taken to
evaluate the representative radial velocity of the knot used to derived the full
spatial velocity (see text). 
\label{vtan}}
\end{figure}

\section{3D kinematics}

The HH~223 emission shows an appreciable undulating  morphology. To better
characterize the velocity field of the emission,  we   searched for the full
spatial velocity and the  inclination angle in the plane of the sky for the HH~223
knots covered by the IFS observations. With this aim, we  combined the proper
motions of  HH~223, derived for the first time in this work, and radial velocities,
derived from the current IFS data.

\subsection{Proper motion determination}

The tangential velocity, \vtan, of the knots was derived from
their proper motions, using three narrow-band H$\alpha$ images of the L723 field,
obtained at the NOT telescope, with
a time baseline of 5 years (see \S 2.2). These images were converted onto a common
reference system using the position of ten common field stars, well distributed
around HH~223, to register the
images. The GEOMAP and GEOTRAN task of IRAF were applied to perform a linear
transformation with six free parameters that take into account relative translation,
rotation and magnification among the frames. The procedure to calculate the proper
motions from multiepoch images has been described in  detail by \citet{Lop05}
and \citet{Ang07} for the cases of HH~110 and HH~30, respectively. In short, it
consists of defining boxes in each frame that include the emission from the
individual knot; then, to
compute the two-dimensional cross-correlation 
function of
the emission
for the two pairs of frames (the 2004 first epoch image and each of
the other two epoch images), and to determine the displacements in $x$ and $y$ coordinates through a
parabolic fit to the peak of the cross-correlation function; 
finally, to perform a
linear regression fit to the displacements of the $x$ and $y$
coordinates as a function of the epoch offset, from the
first epoch (2004). We derived the
proper  motion velocity (\vtan) and the position angle
(PA) for each knot from the slope of the fitted
straight line (and its uncertainty) . The results are displayed in
Fig.~\ref{vtan}, and their values, in columns 2 and 3 of Table~\ref{vtot}.
The errors include the uncertainty in the position of the correlation peak
(derived as explained in \citealp{Lop05}), and the error in the alignment of the
frames.

\subsection{Full spatial velocities}

The IFS data have revealed a complex kinematic structure for the HH~223 emission in
such a way that radial velocity values of the knots show variations depending on
the parcel of the knot emission sampled within the slit (see \eg\ Fig.~\ref{vs79}). 
Note that
the  emission of each knot in  HH~223  extends over several arcsec, both in the
$x$ and $y$ coordinates (see \eg\ Fig.~\ref{cobertura}), being thus improperly sampled 
by a
single long-slit ponting.  Since IFS allows us to achieve a more complete
sampling of the kinematics, we can  assign to each  knot a
representative  radial velocity coming  from the emission enclosed in  the same box
that we defined to  measure the knot proper motion. 
The radial velocity derived
in this way will be used to calculate the full spatial velocity of the knot.
We then
defined the aperture enclosing  each of the boxes A to D (see Fig.~\ref{vtan}), 
and compute the line profiles by  integrating the signal 
over the fibers  inside the aperture. 
We then determined the radial velocity from a
Gaussian fit to the  emission lines 
(see Table~\ref{vrmp}).

The full spatial  velocity (\vtot) and the angle $\phi$ between the knot motion and
the plane of the sky (with positive values of $\phi$ towards the observer) are then
derived from  the H$\alpha$ radial velocity (obtained from IFS data)  and the
proper motions (obtained from H$\alpha$ CCD narrow-band images).  The results are
given in columns 3 and 4 of Table~\ref{vtot}.  Unfortunately, we were only able to
obtain \vtan\ and $\phi$ for the four HH~223 knots covered by the single IFS
pointing. We found that these knots present high full  spatial velocities
($\geq$~100~\kms) and move in a direction close perpendicular to  the plane of the
sky. However, the data suggest that there is some change in the jet  inclination,
which  increases going from the southern knot B (also with the lowest full
velocity) to the  northern knots C and D.

 Apparently, the 3D kinematics of HH~223 (\ie\ the radial velocity structure
and the proper motions) shows close similarities with HH~32 (see, \eg\
\citealp{Sol86}, and \citealp{Cur97}): both HHs are oriented close to the line of
sight and present two-peaked line profiles (note however that in the case of
HH~223, the two-peaked line profiles are not detected at the brightest
knot, HH~223-A). \citet{Rag04} modeled the HH~32 emission 
as a broken-up, large
scale bow-shock with several condensations. The bow-shock
interpretation succeeded in deriving the bow-shock properties, like the
bow-shock velocity (\vshock) and the orientation angle of the outflow in the
plane of the sky ($\phi$), from the components of the proper motion velocity
parallel and perpendicular to the projected outflow axis (\citealp{Rag97}).
However, in spite of the similarities between the HH~32 and 223 structures, the
bow-shock interpretation seems not work so well for HH~223. Assuming a
projected outflow axis at P.A. of 110--115$\degr$ (the P.A. of the
east-west CO bipolar outflow, see \eg\ \citealp{Mor89}; \citealp{Lee02}) 
and from the measured proper
motions, the bow-shock model would give a low shock velocity (\vshock $\leq$ 25 \kms)
and an orientation outflow angle close to the plane of the sky 
($\phi~\leq$ 30$\degr$), which is far from the direction derived using the
observed radial and tangential velocity components 
presented here, even taking into account the uncertainty of
the proper motions due to the uncertainty in the distance to L723.

\section{SUMMARY AND CONCLUSIONS}

We carried out a single-pointing IFS observation of   HH~223, the  optical counterpart of a
larger scale ($\sim$~5.5~arcmin,  $\sim$~0.5~pc at a distance of 300~pc) H$_2$
 outflow, driven by the 
protostellar source VLA~2A, in L723.
Because of  its peculiar and poorly collimated morphology, this target was selected for
the backup list of a science verification program of a new INTEGRAL equalized fiber
bundle. IFS observations have revealed to be the best choice to achieve a reliable
mapping of  the whole spatial emission of such chaotic outflows. The results derived
from IFS data were compared with previous long-slit and CCD narrow-band images looking
for consistency. Unfortunately, because of  the observing time available for our target
in the backup program, it was not possible to acquire an additional  pointing to cover
the fainter HH~223 knots E to F. New results derived from our analysis of the IFS data
are summarized below:

\begin{itemize}
\item
The morphology of the HH~223 emission shown by the line intensity maps built
from the IFS data is in good agreement with the morphology found in the
narrow-band images. One advantage to get line intensity maps from IFS data
is the capability of mapping separately the emission coming from lines that
usually are included into the same narrow-band filter (\eg\ H$\alpha$ and \nii\ 6584
\AA). We obtained in this work a map for each of the emission lines that
revealed the close similarity between the \sii\ and \nii\ emissions and  some
differences between  the spatial brightness distribution 
of the neutral (H$\alpha$) and ionized (\nii\ and \sii) gas.

\item
We traced the spatial distribution of the excitation and the electron density from
line-ratio maps. These maps revealed the complexity of the spatial structure of the
physical conditions. It makes once again IFS mapping  highly appropriate to properly
reveal physical conditions of the emission, instead of derive them from a partial
sampling obtained from long-slit data. As a general trend, the \sii/H$\alpha$ and \nii/H$\alpha$
ratios are consistent with the emission arising from shocks with an intermediate/high
degree of excitation. A relevant result concerning the electron density 
is the inhomogeneities found through the emission (both, through the knots and 
through the low-brightness nebula), which suggests  a clumpy structure 
that cannot be well resolved with the current IFS spatial resolution.

\item
The radial velocity field has been derived, first from the line centroid of a
single-Gaussian fitting to the lines profile and, in addition, from the
cross-correlation technique. Results found from both procedures are fully
consistent. The velocities are highly blueshifted, ranging from --180 to --60
\kms. They present a complex pattern in the spatial distribution, with some 
trend that consists of the more negative (blueshifted) values increasing moving
from southeast to northwest.

\item
The spectra extracted at several positions show clear double-peaked 
line profiles. 
We were able to reproduce the line profiles for these spectra using a model that
included two blueshifted kinematic components. In addition, some trend is found
that consists of the HVC (velocity ranging from --180 to --130 \kms) being more
excited and rarefied than the LVC (velocity ranging from --80 to --40 \kms).
The two velocity components, having quite different
physical properties and spatial distributions,  can be indicative of 
a variability
in the properties of the outflow ejection (\eg\ episodic mass ejection events having
some different velocity and direction). This can be expected when the outflow
source is a binary or a multiple system,  as was proposed by \citet{Car08}.

\item
To search for the reliability of two kinematic components through all the 
emission mapped,
we performed chanel maps both, in H$\alpha$ and
in \sii, for 
two velocity ranges: a HVC (for \vlsr~$\leq$~--90~\kms), and a LVC 
(for \vlsr~$\geq$~--90~\kms).
  We found differences between the spatial distributions 
of the HCV and LVC. The most remarkable differences are found in knots C and B,
where we  
measured offset values from 0.7 to 1.5 arcsec between the positions 
of their  LVC and HVC emission peaks.
Taking into account the spatial resolution of our
data, these offsets are reliable.

\item
We checked the results on the kinematics and physical conditions derived from
long-slit data with those derived from IFS. With this aim, we simulated a long-slit
and extracted spectra with a similar sampling along this slit, positioned as in
long-slit data acquisition. This test lead us to conclude that the physical
conditions and kinematics derived from long-slit spectra are in good agreement with
those derived from IFS spectra that were extracted at the same positions. In contrast, 
we
found that appreciable differences could arise when long-slit data are compared
with IFS spectra that were extracted along slits displaced few  arcsec from the nominal
long-slit position. This test reinforces again the IFS mode to get a more accurate
picture of such as complex outflows.

\item
Comparing the spatial distribution of the gas excitation (Fig.\ref{lineratio})
and kinematics (Fig.\ref{vral}), we can observe a  similar structure of the
maps, which illustrates a relationship between the degree of gas
excitation and the velocity. In particular, the regions having a higher degree
of gas excitation (around knots C and A) coincide with the ones were the 
velocity values are more negative, while emission of lower excitation (located north of
knot B and south of knots C and D) also displays less negative velocity values. 
We cannot appreciate a similar 
relationship when  the maps of the 
electron density and velocity are compared, probably because  the spatial distribution of
the electron density has a pattern more complex than the excitation 

\item
We calculated the full spatial velocity and the
inclination angle of its motion in the plane of the sky for the 
knots mapped with IFS. From proper motion
measurements, using multiepoch narrow-band images, and radial velocities,
derived from IFS, we found that the knots have full spatial velocities higher
than 100 \kms and move close  to the line of sight.
Furthermore, a change in the jet inclination is also outlined.
\end{itemize}

\section*{Acknowledgments}
R.E., R.L. and A.R. were
partially supported by the Spanish MCI grants AYA 2008-06189-C03 and
AYA 2011-30228-C03. 
B.G.-L. acknowledges the support of the Ram\'on y Cajal program, and the 
grants AYA2009-12903 of the Spanish MCI and P/309404 of the Instituto de
Astrof\'{\i}sica de Canarias.
R. L. acknowledges the hospitality of the Instituto de Astrof\'{\i}sica de
Canarias, where part of this work was done.
We thank A. Eff-Darwich for his useful help with the manuscript,  and the
referee, Dr. Raga, for his suggestions and comments.

 \bsp

\label{lastpage}

\end{document}